\documentstyle[a4]{article}
\begin{document}
\begin{center}
\large\bf Comment on
``Additional analytically exact solutions for three-anyons''
and
``Fermion Ground State of Three Particles
in a Harmonic Potential Well and Its Anyon Interpolation''
\end{center}
\begin{center}
\large Stefan Mashkevich
\end{center}
\begin{center}
\it Institute for Theoretical Physics, 252143 Kiev, Ukraine
\end{center}
\begin{center}
(31 December 1996)
\end{center}
\begin{abstract}
The claim put forward in \cite{R95,R96} that the
energies of the ``missing'' states
of three anyons in a harmonic potential
depend linearly on the statistics parameter,
is incorrect because the wave functions proposed
do not satisfy the anyonic interchange conditions.
\end{abstract}

Since a subset of eigenstates of the problem of
three anyons in a harmonic potential was worked out
analytically in \cite{Wu84}
(the energy eigenvalues of those states depend
linearly on the statistics parameter $\alpha$),
much effort has been invested in finding out the
remaining (``missing'') ones.
The results of perturbative and numerical investigations
using different approaches (see \cite{MMO} and references therein)
are all in perfect agreement with each other
and demonstrate, in particular,
that the energies of those ``missing'' states interpolate nonlinearly
between the bosonic and fermionic limits---except for one claim
(not supported by any calculation)
to the effect that the states in question do not exist at all,
violating the continuity of the spectrum \cite{CR92}.

In the papers \cite{R95,R96}, it is argued instead that
the energies of the ``missing'' states depend linearly on $\alpha$;
the continuity of the spectrum is still violated \cite{R95}.
In \cite{R96}, a wave function (7) that turns into the one
of the fermionic ground state for $\alpha=1$ is proposed.
It is then correctly argued that when, say, $z_3=-z_2$
(with $z_j$ the complex particle coordinates)
and $z_1$ describes a small circle around the origin,
an anyon wave function should not change,
because no particles are encircled in the process.
Yet the wave function (7) will in fact change.
The reason is of course that
it contains a piece $\Phi_0 \propto (1+\bar x)^{3-2\alpha}$ (9), with
$x=\frac{z_1 + \eta z_2 + \eta^2 z_3}{z_1 + \eta^2 z_2 + \eta z_3}$
where $\eta=\exp(2\pi{\rm i}/3)$, so that for $z_3=-z_2$ one has
$1+\bar x = \frac{2\bar z_1}{\bar z_1 + {\rm i}\sqrt3 \bar z_2}$;
consequently, rotating $z_1$ around the origin counterclockwise
supplies $\Phi_0$ with a phase factor of $\exp(4{\rm i}\pi\alpha)$
per revolution.
Thus the wave function in question {\it is not a three-anyon
wave function}. The very same argument applies to
the whole class of wave functions of \cite{R95}.

The reasoning in the last paragraph
of Sec.~III of \cite{R96}, where it is claimed that a wave
function with the correct interchange properties can be built
out of (7), amounts essentially to the following:
Given a function $f(\zeta)=\zeta^\alpha$, one wants to turn it
into a single-valued function (i.e., a one that would not change
as a result of $\zeta$ making a revolution around the origin).
There is certainly no way to accomplish this without
making the function discontinuous, which a wave function
cannot be.

To conclude, a striking disagreement of the results of
\cite{R95,R96} with the well-established knowledge comes
from the fact that the issue of multivaluedness of anyon
wave functions has not been properly taken into account
in the papers in question.

\end{document}